# 18-qubit entanglement with photon's three degrees of freedom

Xi-Lin Wang, Yi-Han Luo, He-Liang Huang, Ming-Cheng Chen, Zu-En Su, Chang Liu, Chao Chen, Wei Li, Yu-Qiang Fang, Xiao Jiang, Jun Zhang, Li Li, Nai-Le Liu, Chao-Yang Lu, and Jian-Wei Pan

*Hefei National Laboratory for Physical Sciences at Microscale and Department of Modern Physics, University of Science and Technology of China,*

*& CAS-Alibaba Quantum Computing Laboratory, CAS Centre for Excellence in Quantum Information and Quantum Physics, University of Science and Technology of China, Shanghai 201315, China*

**A central theme in quantum information science is to coherently control an increasing number of quantum particles as well as their internal and external degrees of freedom (DoFs), meanwhile maintaining a high level of coherence. The ability to create and verify multiparticle entanglement with individual control and measurement of each qubit serves as an important benchmark for quantum technologies. To this end, genuine multipartite entanglement have been reported up to 14 trapped ions[1], 10 photons[2,3], and 10 superconducting qubits[4]. Here, we experimentally demonstrate an 18-qubit Greenberger-Horne-Zeilinger (GHZ)[5] entanglement by simultaneous exploiting three different DoFs of six photons, including their paths, polarization, and orbital angular momentum (OAM). We develop high-stability interferometers for reversible quantum logic operations between the photon's different DoFs with precision and efficiencies close to unity, enabling simultaneous readout of $2^{18}$=262,144 outcome combinations of the 18-qubit state. A state fidelity of $0.708 \pm 0.016$ is measured, confirming the genuine entanglement of all the 18 qubits.**

Quantum information is encoded by different states in certain DoFs of a physical system. For example, the quantum information of a single photon can be encoded not only in its polarization[6,7], but also in its time[8], OAM[9], and spatial modes[10]. The simultaneous entanglement with multiple DoFs—known as hyper-entanglement[11]—offers an efficient route to increasing the number of entangled qubits[12,13], and enabled enhanced violations of local realism[14,15], quantum super-dense coding[16], simplified quantum logic gates[17], and teleportation of multiple DoFs of a single photon[18].

Previous experiments have demonstrated hyper-entangled states of two photons in the form of product states of Bell states[12], and fully entangled GHZ states with up to five photons and two DoFs[13]. However, it remained a technological challenge for the multi-photon experiments to go beyond two DoFs. To this end, we develop methods that allow not only scalable creations of hyper-entanglement of multiple photons with three DoFs, but also reversible conversion and simultaneous measurement of multiple DoFs with near-unity precision and efficiency. With these new techniques, we are able to demonstrate and confirm 18-qubit maximal entanglement in GHZ state—the largest Schrödinger cat-like state so far—by manipulating the polarization, spatial modes, and OAM of six photons.

We start by producing polarization-entangled six-photon GHZ states[19,20]. Three pairs of entangled photons are generated by beamlike type-II spontaneous parametric down-conversion (see Fig. 1a) where the signal-idler photon pairs are emitted as two separate circular beams, favorable for being collected into single-mode fiber[2]. The geometry of the down-conversion crystal, where a half-wave plate is sandwiched between two 2-mm-thick β-barium borates, ensures that the obtained photons pairs are polarization entangled[2] in the form of $|\psi^2\rangle = (|H\rangle|V\rangle - |V\rangle|H\rangle)/\sqrt{2}$, where $H$ ($V$) denotes the horizontal (vertical) polarization. The fidelities of the three pairs of entangled photons are measured to be on average $0.98 \pm 0.01$.

Next, we combine photons 2 and 4 on a polarization beam splitter[21] (PBS), and combine one of its output with photon 6 on another PBS (see Fig. 1a). The PBSs transmit *H* and reflect *V*. Fine adjustments of the delays between the different paths are made so that the photons arrive at the PBSs simultaneously. All the six photons are coupled into single-mode fibers and filtered by 3-nm interference filters[7]. Upon detecting one and only one photon in each output, the six photons are projected into the GHZ state in the form of $|\psi^6\rangle = (|H\rangle^{\otimes 6} - |V\rangle^{\otimes 6})/\sqrt{2}$. We obtain a six-fold coincidence count rate of ~0.2 Hz in our experiment.

Thus far, only one DoF of the photons is used. The information-carrying capacity of the photons can be vastly expanded by exploiting other DoFs, including their spatial modes and OAM. To entangling the other DoFs, we apply deterministic quantum logic gates on the single photon's polarization and the other DoFs. Experimentally, we first pass each single photon through a PBS which splits the photon into two paths denoted as up (*U*) and down (*D*) according to its polarization *H* and *V*, respectively. This process can be seen as a controlled-NOT (CNOT) gate where the polarization acts as the control qubit and the path acts as the target qubit, which transforms an arbitrary unknown input single photon in the state $\alpha|H\rangle + \beta|V\rangle$ to a polarization-path hyper-entangled state $\alpha|H\rangle|U\rangle + \beta|V\rangle|D\rangle$. Finally, we encode and entangle the OAM qubit to the photons. Inserting two spiral phase plates (SPPs)[22] in both paths transforms the photon in the *U* and *D* paths into right-handed and left-handed OAM of $+\hbar$ and $-\hbar$ which we denote as $|R\rangle$ and $|L\rangle$, respectively. Each photon is thus prepared in a hyper-entangled state in the form of $\alpha|H\rangle|U\rangle|R\rangle + \beta|V\rangle|D\rangle|L\rangle$. By doing so, starting from the six-photon polarization-entangled GHZ state (Fig. 1a), we arrive at a hyper-entangled 18-qubit GHZ state in the form of $|\psi^{18}\rangle = (|0\rangle^{\otimes 18} - |1\rangle^{\otimes 18})/\sqrt{2}$, where for simplification we denote $|H\rangle$, $|U\rangle$ and $|R\rangle$ as logic $|0\rangle$, and $|V\rangle$, $|D\rangle$ and $|L\rangle$ as logic $|1\rangle$.

The measurement of the 18 individual qubits that expand an effective Hilbert space to 262,144 dimension, and the verification of their multipartite full entanglement, can be technologically more difficult than creating itself. All of the 18 qubits encoded in the three DoFs are to be measured both in the computational base ($|0\rangle, |1\rangle$) and in the superposition base $(|0\rangle \pm e^{i\theta}|1\rangle)/\sqrt{2}$ ($0 \leq \theta \leq \pi$). It is necessary to independently read out one DoF without disturbing any other. The measurements are designed sequentially in three steps.

First, the spatial-mode qubit is measured using a closed or open Mach-Zehnder interferometer, with or without the second 50/50 beam splitter (see Fig. 1c). The open configuration is used to measure the ($|0\rangle, |1\rangle$) base directly. The closed configuration, together with a small-angle prism that adjusts the phase between the two paths, is used to measure the $(|0\rangle \pm e^{i\theta}|1\rangle)/\sqrt{2}$ base. The two outputs (labelled as yellow circles in Fig. 1c) of the open/closed beam splitter correspond to the two orthogonal projection results. In such measurements, the interferometers must be subwavelength stable. We design the beam splitters such that the output modes are parallel and displaced by only 6 mm (see Fig. 1c), and the beam splitters are glued on a glass plate (see Fig. 1f), making the setup insensitive to temperature fluctuations and mechanical vibrations. The current work used six such interferometers, which can remain stable for at least 72 hours with observed visibilities exceeding 99.4%.

The second step is to perform polarization measurement. As shown in Fig. 1d, one of the output from the spatial measurement passes through a quarter-wave plate (QWP), a half-wave-plate (HWP) and a PBS. By adjusting the QWP and HWP at angles of ($0°$, $0°$) and ($45°$, $22.5° - \theta/4$) with respect to the vertical axis, the measurement bases for the polarization states are set for the ($|0\rangle, |1\rangle$) base and the $(|0\rangle \pm e^{i\theta}|1\rangle)/\sqrt{2}$ base respectively. After the PBS, the transmitted or reflected spatial modes corresponds to two orthogonal projection outcomes.

The last step is the readout of the OAM, which, unlike the polarization, was known difficult to be measured with high efficiency and two-channel output simultaneously [9,12,16,23,24]. Our method here is to deterministically map the OAM qubit to the polarization through two consecutive CNOT gates between the two DoFs that together form a quantum swap gate (see the inset of Fig. 1e). In the first CNOT gate, the OAM acts as the control qubit and the polarization acts as the target qubit, which converts the initial state $(\alpha|R\rangle+\beta|L\rangle)|H\rangle$ to an entangled state $\alpha|R\rangle|H\rangle+\beta|L\rangle|V\rangle$. In our experiment, this is achieved using an interferometer which consists of two double-PBSs and two Dove prisms as shown in Fig. 1e (see Methods). In the second CNOT gate, the polarization acts the control qubit and the OAM is the target qubit, transforming the state $\alpha|R\rangle|H\rangle+\beta|L\rangle|V\rangle$ to $(\alpha|H\rangle+\beta|V\rangle)|R\rangle$. Thus, the difficult-to-measure OAM information is coherently transferred to the polarization, which can be conveniently and efficiently readout. Thus, for each single photon carrying three DoFs, the measurement setup can give eight possible outcomes.

Finally, the OAM mode $|R\rangle$ is converted back to the fundamental Gaussian mode (denoted as $|G\rangle$) for efficient coupling into single-mode fibers. This task, together with the second CNOT gate, is completed using one element called $q$-plate[25]. It is an inhomogeneous anisotropic media that couples the polarization with the OAM, transforming $|R\rangle(|H\rangle-i|V\rangle)/\sqrt{2}$ to $|G\rangle(|H\rangle+i|V\rangle)/\sqrt{2}$, and $|L\rangle(|H\rangle+i|V\rangle)/\sqrt{2}$ to $|G\rangle(|H\rangle-i|V\rangle)/\sqrt{2}$, respectively (see Methods). We develop an integrated design for the OAM-to-polarization converter (see Fig. 1h) such that the 24 interferometers used in our work achieve an average visibility of 99.6%, keeping stable for over 72 hours (see Fig. 1g). Using this method, the overall efficiency of the OAM-to-polarization converter is 92%.

The complete experimental setup for creating and measuring the 18-qubit GHZ state is shown in Fig. S1, which includes 30 single-photon interferometers in total. The outputs are detected by 48 single-photon detectors and a complete set of 262,144 combinations can be simultaneously recorded by a coincidence counting system.

To demonstrate the full entanglement among the three DoFs of the $N$-qubit GHZ state, we first simultaneously measure all the qubits along the base of $(|0\rangle \pm e^{i\theta}|1\rangle)/\sqrt{2}$ ($0 \leq \theta \leq \pi$). These measurements give rise to the experimentally estimated expectation values of the observable $M_\theta^{\otimes N} = (\cos\theta\sigma_x + \sin\theta\sigma_y)^{\otimes N}$. For the GHZ states, the expectation value of $M_\theta^{\otimes N}$ in theory fulfills $\langle M_\theta^{\otimes N}\rangle = \cos(N\theta)$, indicating an $N\theta$ oscillation behavior for the expectation value resulting from the collective response to the phase change of all the $N$ entangled qubits. We test such behavior with single-photon polarization state (Fig. 2a), and compare it to three-DoF-encoded GHZ states with one photon (Fig. 2b), four photons (Fig. 2c), and six photons (Fig. 2d), where the phase $\theta$ ramps continuously from 0 to $\pi$. The data are fitted to sinusoidal fringes that show an $N$-times increase in the oscillatory frequencies for the $N$-qubit GHZ states, highlighting the potential of the hyper-entangled states for super-resolving phase measurements[26].

The coherence of the 18-qubit GHZ state, which is defined by the off-diagonal element of its density matrix and reflects the coherent superposition between the $|0\rangle^{\otimes 18}$ and $|1\rangle^{\otimes 18}$ component of the GHZ state, can be calculated by

$$\langle C^{18}\rangle = \frac{1}{18}\sum_{k=0}^{17}(-1)^{k+1}\langle M_{(k\pi/18)}^{\otimes 18}\rangle.$$

From the data shown in Fig. 2d, the coherence is calculated to be $0.602 \pm 0.019$.

For a more detailed characterization of the experimentally created 18-qubit state, we further make measurements at the $|0\rangle/|1\rangle$ base. For an ideal GHZ state, the correct terms in this base should be $|0\rangle^{\otimes 18}$ and $|1\rangle^{\otimes 18}$ only. Our collected data for the $2^{18}$ combinations is plotted in Fig. 2a. By comparing the registered coincidence counts, it shows that the $|0\rangle^{\otimes 18}$ and $|1\rangle^{\otimes 18}$ terms dominate the overall events, with a signal-to-noise ratio (defined as the ratio of the average of the two desired components to that of the remaining non-desired ones) of $5.7 \times 10^5 : 1$. Thus we can calculate the population of the $|0\rangle^{\otimes 18}$ and $|1\rangle^{\otimes 18}$ terms is $0.814 \pm 0.026$. By analyzing the undesired components in Fig. 2e, we estimate that ~11.3% noise is contributed from the double pair emission of parametric down-conversion noise and the remaining ~7.3% is from bit-flip error due to the imperfection of the optical elements such as the PBS and interferometers.

The state fidelity, which is defined as the overlap of the experimentally generated state with the ideal one: $F(\psi^{18}) = \langle \psi^{18} | \rho_{exp} | \psi^{18} \rangle = Tr(\rho_{ideal} \rho_{exp})$. The fidelity can be directly calculated by the average of the expectation values of the population and coherence. From the experimental results as shown in Fig. 2, the fidelity of generated entangled 18-qubit GHZ state is calculated to be $0.708 \pm 0.016$. The notion of genuine multipartite entanglement characterizes whether generation of the state requires interaction of all parties, distinguishing the experimentally produced state from any incompletely entangled state. For the GHZ states, it is sufficient for the presence of genuine multipartite entanglement[27] if their fidelities exceed the threshold of 0.5. Thus, with high statistical significance (>13σ), our experiment confirms the genuine 18-qubit entanglement, the largest entangled state demonstrated so far with individual control of each qubit.

In summary, we have developed methods for precise and efficient quantum logic operations on multiple DoFs of multiple photons, and generated and verified the full entanglement among 18 qubits. Our work demonstrates that combining multi-particle entanglement with multiple internal and external DoFs can provide an efficient route to increase the number of effective qubits. For instance, with the same parametric down-conversion source, if only one DoF (polarization) is exploited, an 18-photon GHZ state would have a count rate of $2.6 \times 10^{-15}$ Hz. Exploiting three DoFs, our hyper-entangled 18-qubit GHZ state is ~13 orders of magnitude more efficient than the single-DoF 18-photon GHZ state. Our work creates a new platform for optical quantum information processing with multiple DoFs. The ability to coherently control 18 qubits enables experimental access to previously unexplored regimes, for example, the realization of the surface code[28] and the Raussendorf–Harrington–Goyal code[29]. It will be interesting in future work to exploit the high quantum numbers[30-32] of the OAM or path to create new type of entanglement[33].

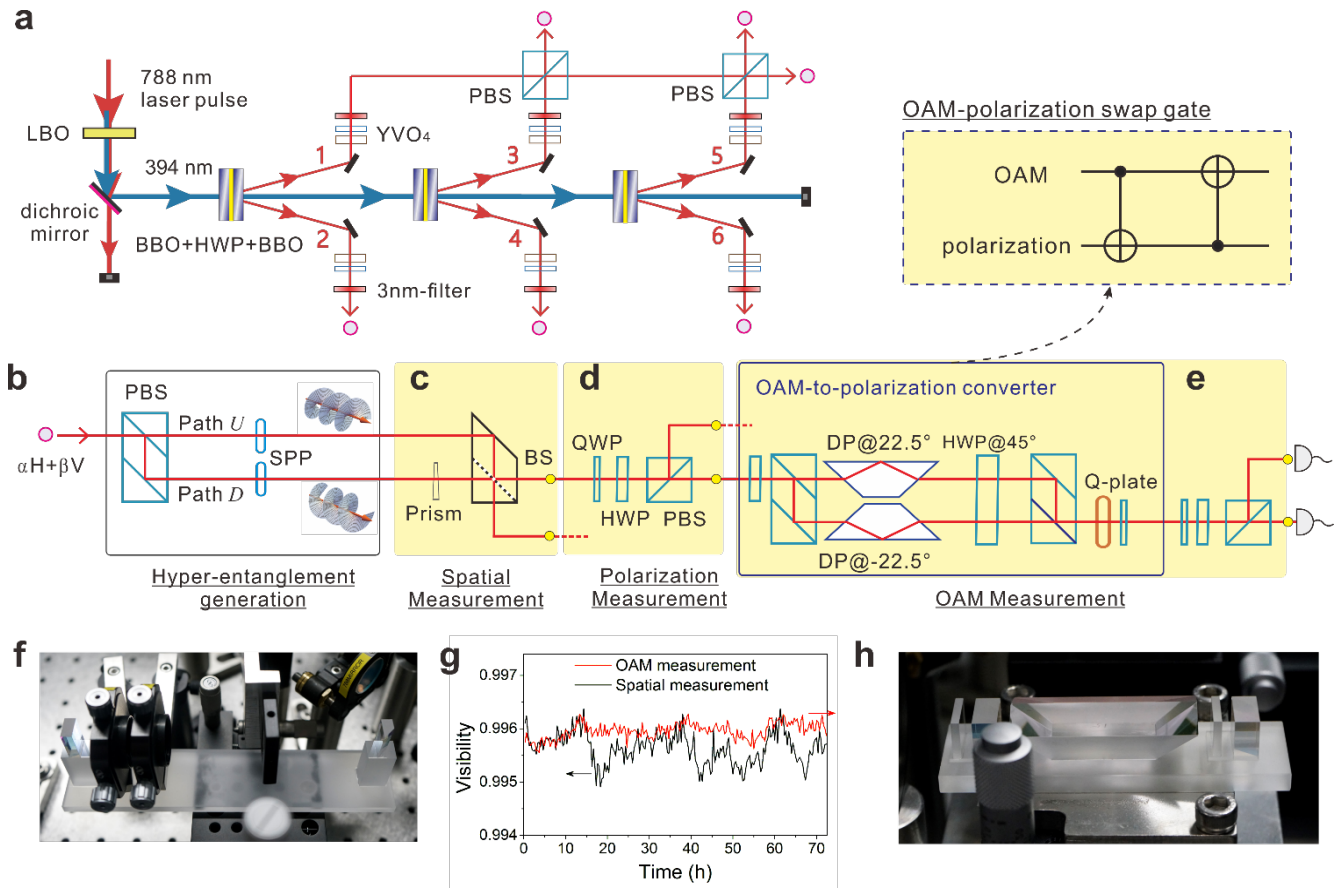

**Figure 1**: Scheme and experimental setup for creating and verifying 18-qubit GHZ state consisting of six photons and three degrees of freedom. **a.** The generation of six-photon polarization-entangled GHZ state. An ultrafast laser with a central wavelength of 788 nm, a pulse duration of 120 fs and a repetition rate of 76 MHz is focused on a lithium triborate (LBO) and up-converted to 394 nm. The ultraviolet laser is focused on three custom-designed sandwich-like nonlinear crystals, each consists of two 2-mm-thick β-barium borates (BBOs) and one half-wave plate (HWP), to produce three pairs of entangled photons. In each output, two pieces of $YVO_4$ crystals with different thickness and orientation are used for spatial and temporal compensation for the birefringence effects. The three pairs of entangled photons are combined on two polarizing beam splitters (PBSs) to generate a six-photon polarization-entangled GHZ state. **b.** For each single photon, it is sent through a double PBS and two spiral phase plates (SPPs) to be prepared in a single-photon three-qubit state. **c.** The measurement of the spatial qubit with closed (dash line) or open (without the dash line) interferometric configuration. **d**. Polarization measurement. **e**. High-efficiency and dual-channel OAM readout by coherently convert the OAM to polarization by a swap gate (inset). **f**. Photo of the actual setup used in **b** and **c**. By vertical translation, it is convenient to switch between open and close **g**. Real-time monitoring the visibilities in the spatial (**f**) and OAM (**h**) measurements. **h**. Photo of the actual setup used in **e**. DP: Dove prism.

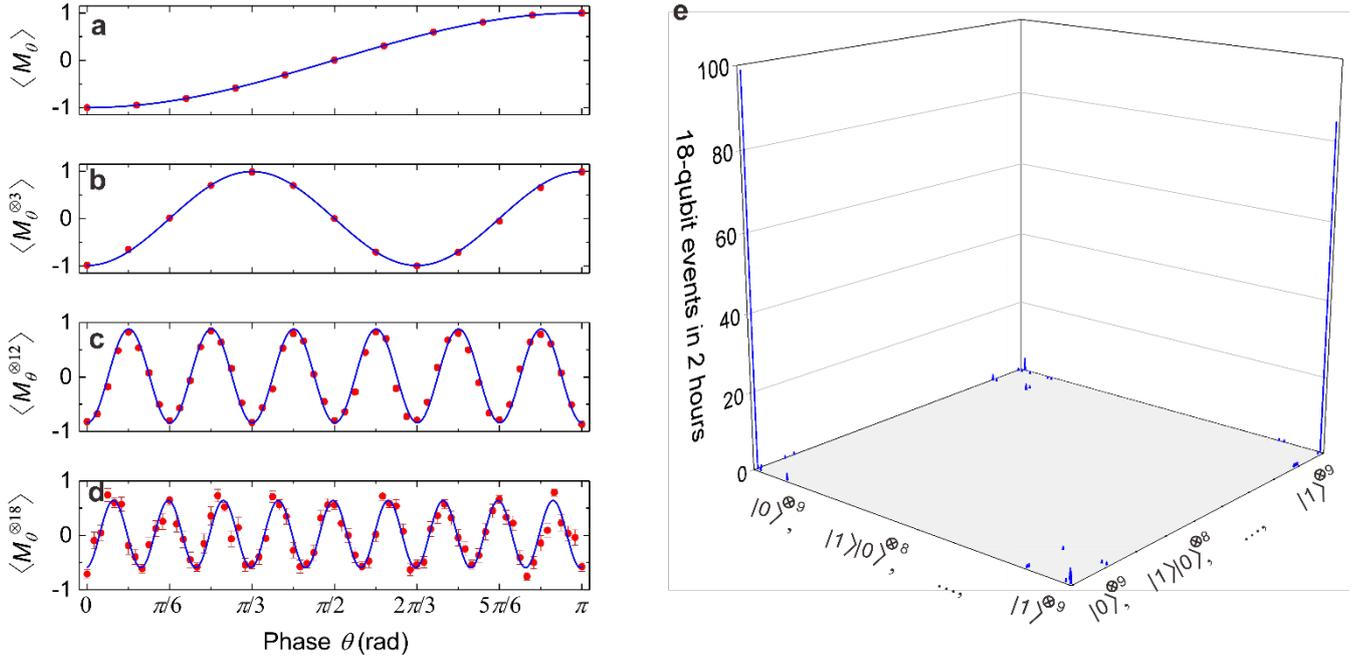

**Figure 2**: Experimental data of 18-qubit GHZ entanglement. The **a**, $N = 1$, **b**, $N = 3$, **c**, $N = 12$, and **d**, $N = 18$ qubits are measured in the superposition base $(|0\rangle \pm e^{i\theta}|1\rangle)/\sqrt{2}$. Each of the $N$-qubit events corresponds to the observation of an eigenstate of the observable $M_\theta^{\otimes N} = (\cos\theta\sigma_x + \sin\theta\sigma_y)^{\otimes N}$ with eigenvalue of $v_s = +1$ or $v_s = -1$. The expectation values $\langle M_\theta^{\otimes N} \rangle$ can be calculated by $\langle M_\theta^{\otimes N} \rangle = \sum_{s=1}^{2^N} p_s v_s$, where $p_s$ $(s = 1,...,2^N)$ is the relative probability of the $N$-qubit detection events. The x-axis denotes the phase shift $\theta$ between $|0\rangle$ and $|1\rangle$, and the y-axis is the experimentally obtained $\langle M_\theta^{\otimes N} \rangle$. Error bars indicate one standard deviation and are calculated by the experimentally detected $N$-qubit events with the propagated Poissonian counting statistics. In **a**-**c**, the error bars are smaller than the data points. **e,** 18-qubit events in the computational $|0\rangle/|1\rangle$ basis, accumulated for two hours, are displayed in a $512 \times 512 = 262,144$ dimensional matrix.

## Methods

**OAM-polarization CNOT gate:** The interferometer is designed to convert an OAM state $(\alpha|R\rangle+\beta|L\rangle)|H\rangle$ to an OAM-polarization entangled state $\alpha|R\rangle|H\rangle+\beta|L\rangle|V\rangle$. To do so, the state is first transformed to the state $(\alpha|R\rangle+\beta|L\rangle)(|H\rangle+|V\rangle)/\sqrt{2}$ by a HWP at 22.5°. Second, the photon passes through the interferometer composing of two PBSs and two Dove prisms. The first PBS splits the input photon into up and down paths according to the polarization state of *H* and *V*. Passing through a Dove prism rotated at an angle of $\theta$, an OAM state $|R\rangle(|L\rangle)$ will pick up a phase of $e^{-i2\theta}(e^{i2\theta})$. Thus, the two Dove prisms rotated at 22.5° or -22.5° respectively in the up and down paths transform the state $(\alpha|R\rangle+\beta|L\rangle)|H\rangle/\sqrt{2}$ and $(\alpha|R\rangle+\beta|L\rangle)|V\rangle/\sqrt{2}$ into $(\alpha|R\rangle e^{-i\pi/4}+\beta|L\rangle e^{i\pi/4})|H\rangle/\sqrt{2}$ and $(\alpha|R\rangle e^{i\pi/4}+\beta|L\rangle e^{-i\pi/4})|V\rangle/\sqrt{2}$. A HWP at 45° switches the polarization and enables the recombination of the *H* and *V* component at the second PBS. The output from the interferometer reads

$$e^{i\pi/4}\alpha|R\rangle(|H\rangle-i|V\rangle)/\sqrt{2}+e^{-i\pi/4}\beta|L\rangle(|H\rangle+i|V\rangle)/\sqrt{2}.$$

With a QWP rotated at -45°, the state is converted into $\alpha|R\rangle|H\rangle+\beta|L\rangle|V\rangle$, which is the exact entangled state we aim for.

**The *q*-plate:** The *q*-plate is a spatial-variant wave plate with inhomogeneous optical axis distribution. Here, the optical axis of the *q*-plate is orientated in the direction $\phi/2$, where $\phi$ is the azimuthal angle in the polar coordinate system. A circularly polarized photon in the state of $(|H\rangle+i|V\rangle)/\sqrt{2}$ and $(|H\rangle-i|V\rangle)/\sqrt{2}$ with zero OAM is converted to the state of $e^{i\phi}(|H\rangle-i|V\rangle)/\sqrt{2}$ and $e^{-i\phi}(|H\rangle+i|V\rangle)/\sqrt{2}$, with OAM of $+\hbar$ and $-\hbar$ in the respectively. Therfore, by passing through the *q*-plate, the OAM-polarization entangled state $\alpha|R\rangle|H\rangle+\beta|L\rangle|V\rangle$ outputting from the interferometer for OAM measurement, is transferred to $(\alpha|H\rangle+\beta|V\rangle)|G\rangle$ (where $|G\rangle$ denotes the fundamental Gaussian mode for efficiently coupling into single-mode fibers) together with two QWPs following by the rules as listed below:

$$|R\rangle|H\rangle \xrightarrow{\text{QWP}} |R\rangle(|H\rangle-i|V\rangle)/\sqrt{2} \xrightarrow{q\text{-plate}} |G\rangle(|H\rangle+i|V\rangle)/\sqrt{2} \xrightarrow{\text{QWP}} |G\rangle|H\rangle,$$

$$|L\rangle|V\rangle \xrightarrow{\text{QWP}} |L\rangle(|H\rangle+i|V\rangle)/\sqrt{2} \xrightarrow{q\text{-plate}} |G\rangle(|H\rangle-i|V\rangle)/\sqrt{2} \xrightarrow{\text{QWP}} |G\rangle|V\rangle.$$